\documentstyle[12pt,fleqn]{article}
\def\beq{\begin{equation}}
\def\eeq{\end{equation}}

\begin{document}
\vspace*{15mm}
\begin{center}
{\Large Chaotic Evolution in Quantum Mechanics} \\[2cm]
Asher Peres$^*$ \\[7mm]
{\sl Department of Physics, Technion---Israel Institute of
Technology, 32 000 Haifa, Israel}\vfill

(10 August 1995) \end{center}\vfill

\noindent{\bf Abstract}\bigskip

A quantum system is described, whose wave function has a
complexity which increases exponentially with time. Namely, for any
fixed orthonormal basis, the number of components required for an
accurate representation of the wave function increases
exponentially.\vfill

\vfill

\noindent $^*$\,Electronic address: peres@photon.technion.ac.il

\vfill\newpage

This Letter describes a quantum system whose representation becomes
increasingly complex with the passage of time. This behavior, which is
well known as a generic property of {\it classical\/} Hamiltonian
systems, is commonly called ``chaos.'' For example, an initially smooth
Liouville density becomes more and more convoluted as time passes, and
it may form intricate shapes with exceedingly thin and long
protuberances. However, quantum wave functions usually have a milder
behavior [1]. In particular, quantum systems with discrete spectra can
be represented, with arbitrary accuracy, by a {\it finite\/} number of
eigenfunctions which are periodic in time. Their wave function thus is
almost periodic. Its {\it computational complexity\/} [2, 3] does not
increase as time passes. (Some authors use other criteria for
complexity, such as the visual appearance of wave functions, but these
seem rather subjective.)

In this Letter, I describe a quantum system whose wave function becomes
increasingly complex with time, just as a Liouville density would do in
classical mechanics. The long range evolution of the wave function is
effectively unpredictable with finite computing resources. This is a
genuine example of {\it quantum chaos\/}. I have no pretense that the
abstract model discussed here represents, even approximately, a real
physical object. However, this is a formal proof that, contrary to some
claims, quantum mechanics is compatible with the existence of chaos.

Such a chaotic quantum system can be constructed for any classical
area-preserving map. As a concrete example, I shall choose a map
consisting of alternating twists and turns of a unit sphere. That map
has a quantum version, called ``kicked top'' [4], with a discrete,
finite-dimensional quasi-energy spectrum. Therefore, unlike the
classical map, the quantum kicked top is not truly chaotic. However, a
classical dynamical system may have several, different quantum versions.
Another quantization of the same twist and turn (T\&T) map is presented
here, which is just as chaotic as the original classical map.

The latter is defined as follows: consider the unit sphere,
$x^2+y^2+z^2=1$. Each step of the T\&T map consists of a twist by an
angle $a$ around the $z$-axis (namely, every $xy$ plane turns by an
angle~$az$), followed by a 90$^\circ$ rigid rotation around the
$y$-axis. The result is

\beq\begin{array}{l}
 x'=z,\smallskip \\
 y'=x\,\sin(az)+y\,\cos(az),\smallskip \\
 z'=-x\,\cos(az)+y\,\sin(az).
\end{array} \label{TT} \eeq
This map is obviously area preserving. For low values of~$a$, most
classical orbits are regular (that is, they are quasi-periodic). As $a$
increases, so does the fraction of chaotic orbits, until for $a=3$ most
of the sphere is visited by a single chaotic orbit (all numerical
calculations below refer to the case $a=3$).

The ``kicked top'' [4] is a mechanical system (classical or quantal)
which mimics the above geometrical map. In the classical case, the top
has a 3-dimensional generalized phase space [5], with canonical
variables $J_x,\ J_y$, and $J_z$, satisfying Poisson brackets
$\{J_x,J_y\}=J_z$, and cyclic permutations. The mapping

\beq\begin{array}{l}
 J_x'=J_z,\smallskip \\
 J_y'=J_x\,\sin(aJ_z/J)+J_y\,\cos(aJ_z/J),\smallskip \\
 J_z'=-J_x\,\cos(aJ_z/J)+J_y\,\sin(aJ_z/J),
\end{array} \label{JJ} \eeq
is a canonical transformation [6] which leaves $J^2$ invariant. The
classical values of $J_x/J$, $J_y/J$, and $J_z/J$, lie on a unit sphere,
and therefore they transform just as $x,\ y$ and $z$ in Eq.~(1).

A natural way of quantizing these equations is to replace the classical
variables $J_k$ by the corresponding quantum operators. Equation (2)
then becomes a quantum map, generated by the {\it unitary\/}
transformation

\beq U=\exp(-i\pi J_y/2\hbar)\,\exp(-iaJ_z^2/2J\hbar).
\label{U} \eeq
For any eigenstate of $J$, this $U$ is a matrix of order $(2j+1)$, with
a discrete spectrum. The evolution is multiply periodic, and there is no
chaos analogous to that of the classical map: computing the wave
function for long times is not more difficult than for short times.
(There still is hypersensitivity to small perturbations [7], but this is
not ``chaos'' in the classical sense.)

We can also consider quantum systems which are not restricted to a
particular value of~$j$. Their Hilbert space is spanned by the spherical
harmonics $Y^m_j(\theta,\phi)$, with angles $\theta$ and~$\phi$ related
to the cartesian coordinates in Eq.~(\ref{TT}) in the usual way:
$x=\sin\theta\,\cos\phi$, etc. The unitary evolution is still generated
by Eq.~(\ref{U}), where it is now understood that the various $J_k$ have
to be written as block-diagonal matrices, with blocks of order $(2j+1)$.
In that case, it is also possible to write the wave function as

\beq \psi(\theta,\phi)=\sum_{jm} C_{jm}\,Y^m_j(\theta,\phi),
  \label{expsi} \eeq
but its components with different $j$ are never mixed by the $U$ matrix.
They evolve independently of each other. There still is no chaos: the
cost of computing the final state does not increase with the number of
steps. We thus see that the quantum kicked top is not a faithful replica
of the classical T\&T map. The reason for this failure of the
correspondence principle is explained below, and an alternative
quantization method will be proposed.

Consider the Hamiltonian which generates the unitary map
(\ref{U}):

\beq H=aJ_z^2/2J+(\pi J_y/2)\,\sum_n\delta(t-n),
 \label{Hk}\eeq
where the unit of time is the duration of one step. In this Hamiltonian,
the twist is continuous, and the rotation proceeds by kicks (the
opposite choice is also possible). If $\psi$ is represented by a set of
$C_{jm}$ coefficients as in Eq.~(\ref{expsi}), the $J_k$ in (\ref{Hk})
are block-diagonal, as explained above. On the other hand we may as well
use the ``coordinate basis'' and directly write $\psi(\theta,\phi)$
without expanding into spherical harmonics, which is closer to the
spirit of the classical T\&T map. We then have
\,$J_z=-i\hbar\partial/\partial\phi$, and more complicated differential
operators for $J_x$ and $J_y$ [8]. In particular,

\beq
 J=-i\hbar\left[{1\over\sin\theta}\,{\partial\over\partial\theta}\,
 \left(\sin\theta\,{\partial\over\partial\theta}\right)+
 {1\over\sin^2\theta}\,{\partial^2\over\partial\phi^2}
 \right]^{1/2}. \eeq

When we were in a finite-dimensional Hilbert space, it was natural
(indeed unavoidable) to write the twist generator as $J_z^2/2J$.
However, for arbitray $\psi(\theta,\phi)$, a more natural expression
for the twist generator is $J_z\cos\theta$. Let us therefore replace
(\ref{Hk}) by

\beq H=aJ_z\cos\theta+(\pi J_y/2)\,\sum_n\delta(t-n).
 \label{HTT}\eeq
This is a Hermitian operator which does not commute with $J^2$, so that
$J^2$ is not conserved. (In other words, $J_z\cos\theta$ has
nonvanishing matrix elements between $Y_j^m$ with different $j$, and
cannot be written solely in terms of the $J_k$ matrices.) Instead of
(\ref{U}) we now have

\beq U=\exp(-i\pi J_y/2\hbar)\,\exp(-iaJ_z\cos\theta/\hbar).
\label{UTT} \eeq

The crucial difference is that the spectra of the new $H$ and $U$
include {\it continuous\/} parts, and therefore permit the existence of
true chaos. This can be shown as follows. Returning to the classical
T\&T map (1), let us consider, instead of individual points, a mass
density $\rho(\theta,\phi)$ spread on the unit sphere. We may even
consider {\it complex\/} densities, if we wish. Let us further assume
that the infinitesimal mass, $\rho(\theta,\phi)d\Omega$, attached to the
area element $d\Omega=\sin\theta\,d\theta\,d\phi$, is conserved by the
T\&T map. It thus behaves as an incompressible fluid, or as a Liouville
density would do while moving in phase space.

As the map is area preserving, namely $d\Omega'=d\Omega$, we have, at
each step,

\beq  \rho'(\theta,'\phi')=\rho(\theta,\phi). \eeq
This implies that the map is {\it unitary\/}: for any two densities
$\rho_1$ and $\rho_2$,

\beq \int\rho'_1(\theta',\phi')^*\,\rho'_2(\theta',\phi')\;d\Omega'
 =\int\rho_1(\theta,\phi)^*\,\rho_2(\theta,\phi)\;d\Omega. \eeq
Such a unitarity property was proved long ago by Koopman [9] for
Liouville densities in phase space. There is a large body of knowledge
on the ergodicity and mixing properties of Liouville functions [10]: In
the generic non-integrable case, the Liouvillian has a {\it
continuous\/} spectrum, in which an infinite number of discrete lines
may be embedded. If the dynamical system has a finite measure, the
unitary operator $U$ has at least one eigenvalue equal to 1,
corresponding to equilibrium. Moreover, it can be proved~[10] that, if
the system is ergodic, but not mixing, that eigenvalue is nondegenerate,
and all the other eigenvalues of $U$ form a subgroup of the circle
group. On the other hand, for a mixing system, which also has a single
nondegenerate eigenvalue 1, the rest of the spectrum is absolutely
continuous. A generic dynamical system may have some regions of phase
space which are subject to mixing, others which are only ergodic, and
still others which are not even ergodic. Such a system is called
``decomposable''~[11]. In that case, the spectrum of $U$ is continuous,
with an infinite number of discrete lines embedded in it. In particular,
the eigenvalue 1, and possibly others, are degenerate.

Now, it is easily seen that the quantum wave function
$\psi(\theta,\phi)$ behaves, under the unitary transformation $U$ in
Eq.~(\ref{UTT}), exactly as the fictitious incompressible mass density
described above.  This is obvious for the rotation operator, $\exp(-i\pi
J_y/2\hbar)$, which performs a rigid rotation of $\psi$ around the
$y$-axis. For the twist, $\exp(-iaJ_z\cos\theta/\hbar)$, we have

\beq  \psi(\theta,\phi)\to\psi'(\theta,\phi)=
  \psi(\theta,\phi-a\,\cos\theta). \label{twist}\eeq
That is, the wave function $\psi$ moves on the surface of the unit
sphere exactly as the classical points did, and it remains normalized,
by virtue of the unitarity of the mapping. The essential difference
between this new quantum system and the former kicked top is that the
new Hamiltonian (which formally behaves as a classical Liouville
operator) has a partly continuous spectrum. Therefore the evolution of
$\psi(\theta,\phi)$ cannot be represented, even approximately, by a
finite number of terms.

Let us examine this evolution more closely. The rotational part of $U$
is represented, in the $Y_j^m$ basis, by the familiar orthogonal matrix
$U^{(j)}_{mm'}$, namely a block-diagonal matrix with blocks of size
$(2j+1)$ [12, 13]. In the same basis, the twist (\ref{twist}) is also
represented by a unitary transformation [14],

\beq C_{jm}\to C'_{jm}=\sum_l U^{(m)}_{jl}\,C_{lm}, \eeq
where

\beq U^{(m)}_{jl}=\int Y_j^{m*}(\theta,\phi)\,Y_l^m(\theta,\phi)\,
  e^{-ima\cos\theta}\,d\Omega, \eeq
is a band matrix. This transformation leaves $m$ invariant, but
introduces all the $j$ with $j>|m|$ (with exponentially small
coefficients for large $j$).

It is now possible to give a quantitative measure for the complexity
of the wave function. Its information entropy [15],

\beq S=-\sum_{jm}\,|C_{jm}|^2\,\log |C_{jm}|^2, \eeq
has the intuitive meaning that $e^S$ roughly indicates the number of
basis vectors that are appreciably involved in the expansion of $\psi$
into spherical harmonics. This ``entropy'' of course depends of the
choice of the basis (namely, spherical harmonics).  However, it can be
shown [14] that when $S$ becomes large, its value is asymptotically
independent of the choice of the basis, provided that the transformation
between different bases is given by a band matrix (the latter property
holds for any two bases whose definition is not algorithmically
complex).

Finally, let us illustrate the above considerations by a numerical
example. A plot of $S$ versus the number of steps is shown in Fig.~1,
for $a=3$ and for two initial states, $\psi_\pm$, with

\beq C_{11}=\pm\,C_{1,-1}=1/\sqrt{2}, \label{C11}\eeq
respectively, and all other $C_{lm}=0$. It is seen that $S$ increases
roughly linearly (with a small negative second derivative), so that, as
the wave function evolves and becomes more and more complicated, the
effective number of components needed for representing it increases
about exponentially with time. (The calculation of the last step in
Fig.~1 involved all the matrices with $j,\ l,\ m$, up to 500. The next
step would have exceeded the capacity of my workstation, or entailed a
severe loss of accuracy.)\bigskip

Some of the ideas presented here originated during a visit to the
International Center of Theoretical Physics, Trieste, whose hospitality
is gratefully acknowledged. I had interesting discussions with Felix
Izrailev and Dima Shepelyansky. This work was supported by the Gerard
Swope Fund, and the Fund for Encouragement of Research.\clearpage

\frenchspacing\begin{enumerate}
\item H. J. Korsch and M. V. Berry, Physica D {\bf 3}, 627 (1981).
\item C. H. Bennett, ``Dissipation, Information, Computational
 Complexity and the Definition of Organization'' in {\it Emerging
 Syntheses in Science\/}, ed. by D. Pines (Addison-Wesley, Reading, MA,
 1987) pp.~215--231.
\item G. Brassard and P. Bratley, {\it Fundamentals of Algorithmics\/}
 (Prentice-Hall, Englewood Cliffs, 1995) Chapt.~12.
\item F. Haake, M. Ku\'s, and R. Scharf, Zeits. Phys. B {\bf 65}, 381
 (1987).
\item J. L. Martin, Proc. Roy. Soc. (London) A {\bf 251}, 536 (1959).
\item A. Peres, {\it Quantum Theory: Concepts and Methods\/} (Kluwer,
 Dordrecht, 1993) p.~300.
\item A. Peres, ``Instability of Quantum Motion of a Chaotic System'' in
 {\it Quantum Chaos: Proceedings of the Adriatico
 Research Conference on Quantum Chaos\/}, ed. by H. A. Cerdeira, R.
 Ramaswamy, M. C. Gutzwiller, and G.~Casati (World Scientific,
 Singapore, 1991) pp.~73--102.
\item L. I. Schiff, {\it Quantum Mechanics\/} (McGraw Hill, New York,
1968) p.~82.
\item B. O. Koopman, Proc. Nat. Acad. Sci. {\bf 17}, 315 (1931).
\item P. R. Halmos, {\it Lectures on Ergodic Theory\/} (Chelsea, New
 York, 1956) pp. 34--46.
\item V. I. Arnold and A. Avez, {\it Ergodic Problems of Classical
 Mechanics\/} (Benjamin, New York, 1968) p. 17.
\item E. P. Wigner, {\it Group Theory and Its Application to the
 Quantum Mechanics of Atomic Spectra\/} (Academic Press, New York, 1959)
 p.~167.
\item M. Tinkham, {\it Group Theory and Quantum Mechanics\/}
 (McGraw-Hill, New York, 1964) p.~110.
\item A. Peres and D. Terno, ``Evolution of Liouville Density of a
Chaotic System'' (submitted to Phys. Rev. E).
\item F. M. Izrailev, Phys. Lett. A {\bf 134}, 13 (1988); J. Phys. A
{\bf 22}, 865 (1989).
\item J. Ford, ``Chaos: Solving the Unsolvable, Predicting the
Unpredictable!'' in {\it Chaotic Dynamics and Fractals\/} ed. by M. F.
Barnsley and S. G. Demko (Academic Press, New York, 1986) pp.~1--52.
\end{enumerate}\nonfrenchspacing\vfill

\parindent 0mm
{\bf Caption of figure}\bigskip

FIG. 1. \ Growth of $S$ for the two initial states in Eq.~(\ref{C11}).
\end{document}